\documentclass[11pt,a4paper]{article}
\usepackage[left=15mm, top=10mm, right=15mm, bottom=20mm, nohead=true, nofoot=true]{geometry}

\usepackage[utf8]{inputenc}
\usepackage{authblk}
\usepackage{indentfirst}
\usepackage{misccorr}
\usepackage{graphicx}
\usepackage{amsmath}
\usepackage{amssymb}
\usepackage{multicol}
\usepackage{bm}
\usepackage{longtable}
\usepackage{pdflscape}
\usepackage{epstopdf}
\usepackage{multirow}
\usepackage{adjustbox}
\usepackage{amsfonts}
\usepackage{ifthen}
\usepackage{fancyhdr}
\usepackage{setspace}
\usepackage{xcolor}
\usepackage[round,authoryear,comma,sort&compress,numbers]{natbib}
\usepackage[toc,title,page]{appendix}
\usepackage[hidelinks]{hyperref}
\usepackage{url}
\usepackage{caption}
\usepackage{booktabs}

\newcommand{\Msun}{\ensuremath{M_{\odot}}}

\newcommand{\Teff}{\ensuremath{T_{\rm eff}}}

\hypersetup{
    colorlinks=true,
    linkcolor=green,
    citecolor=blue,
    filecolor=magenta,
    urlcolor=cyan,
    pdftitle={Dust in Stellar Populations: Grid-Based and Bayesian Spectral Energy Distribution Fitting of AGB Stars with SEDust and MCM-SEDust},
    pdfpagemode=FullScreen,
}

\fancypagestyle{plain}{\chead{\texttt{\textcolor{brown}{Communications of BAO, Vol. ?, Issue ?, 20??, pp. ???-???}}}}
\title{Dust in Stellar Populations: Grid-Based and Bayesian Spectral Energy Distribution Fitting of AGB Stars with \texttt{SEDust} and \texttt{MCM-SEDust}}
\author[1,2]{Hamidreza Mahani \thanks{mahani@ipm.ir}}
\author[1,2]{Atefeh Javadi}
\author[1]{Amirhossein Rabiee}
\affil[1]{\scriptsize School of Astronomy, Institute for Research in Fundamental Sciences (IPM), P.O. Box 1956836613, Tehran, Iran}
\affil[2]{\scriptsize Iranian National Observatory (INO), Institute for Research in Fundamental Sciences (IPM), P.O. Box 1956836613, Tehran, Iran}

\begin{document}
\pagestyle{empty}
\newpage
\pagestyle{fancy}
\label{firstpage}
\date{}
\maketitle

\begin{abstract}

Dust shapes the spectral energy distributions (SEDs) of stellar populations on every scale, from the circumstellar envelopes of evolved stars to the interstellar medium of galaxies. We review the ingredients of stellar population synthesis (SPS) models: the initial mass function, stellar evolutionary tracks, stellar spectral libraries, and dust models. We then present \texttt{SEDust}, an automated pipeline that finds the best-fitting \texttt{DUSTY} radiative-transfer model for the SEDs of dusty asymptotic giant branch (AGB) stars, and \texttt{MCM-SEDust}, its extension. We discuss how both connect the physics of individual dusty stars to the dust budget of stellar populations.

\end{abstract}


\section{Introduction}
\label{sec:intro}

Stellar population synthesis (SPS) is the main tool for interpreting the integrated light of star clusters and galaxies. Its aim is to recover physical properties such as stellar mass, age, metallicity, star formation history (SFH) and dust content from observed SEDs \citep{Tinsley1968,Conroy2013, Zonoozi19, Mahani21}. The method rests on the idea that a galaxy is a superposition of stars formed at different times, so its light can be modelled by summing the contributions of simple (single-age, single-metallicity) stellar populations (SSPs), weighted by the SFH and modified by dust:
\begin{equation}
  F_\lambda(t) = \int_0^{t} \mathrm{d}t' \int_0^{\infty} \mathrm{d}Z\;
  \psi(t - t')\, P(Z,t')\, S_\lambda\!\left[t', Z\right]\, e^{-\tau_\lambda(t,t')} \;+\; F_\lambda^{\rm dust\,em},
  \label{eq:csp}
\end{equation}
where $\psi$ is the star formation rate, $P(Z,t')$ the metallicity distribution, $S_\lambda$ the SSP luminosity per unit mass, $\tau_\lambda$ the dust optical depth, and $F_\lambda^{\rm dust\,em}$ the re-emitted infrared (IR) radiation. Each SSP is built from four ingredients, discussed below. Uncertainties in any of them propagate into the derived galaxy properties \citep{Conroy2009,Conroy2013}.

\subsection{Initial mass function}
The initial mass function (IMF), $\xi(m)=\mathrm{d}N/\mathrm{d}m$, gives the distribution of stellar masses at birth. It sets the relative numbers of massive stars, which dominate the ultraviolet (UV) output and the production of heavy elements and dust, and of low-mass stars, which dominate the stellar mass and the near-IR light at late times. Commonly adopted forms are the single power law of \citet{Salpeter1955}, $\xi(m)\propto m^{-2.35}$, and the broken power-law and log-normal forms of \citet{Kroupa2001} and \citet{Chabrier2003}, which flatten below $\sim 0.5\,\Msun$. The IMF normalizes the stellar mass-to-light ratio, so a change of IMF rescales the inferred stellar mass. Variations of the IMF with environment remain actively debated \citep{Conroy2013, Mahani21}.

\subsection{Stellar evolutionary tracks and isochrones}
Evolutionary tracks give the luminosity and effective temperature of a star of given initial mass and metallicity as a function of time. Interpolating them yields isochrones, which are the population's Hertzsprung--Russell diagram at a fixed age. Widely used sets include the Padova/PARSEC libraries \citep{Girardi2000,Bressan2012}, the MIST tracks \citep{Choi2016}, and the BaSTI database. The most uncertain phases are the thermally pulsing AGB (TP-AGB), core helium burning, and the effects of rotation and binarity. The TP-AGB is short-lived but very luminous in the near-IR, and it is also where much of the stellar dust is produced \citep{Saremi20, Abdollahi23_IAU, Javadi19, Saremi19, Hashemi19}. \citet{Maraston2005} and \citet{Marigo2008} showed that its treatment changes the predicted near-IR light of intermediate-age populations substantially. The star-by-star dust modelling of AGB stars described below is therefore directly relevant to SPS models \citep{Mahani23, Rostami25, Gholami25, Mahani_Farsi}.

\subsection{Stellar spectral libraries}
To convert each point of an isochrone ($\Teff$, $\log g$, $[{\rm Fe/H}]$) into a spectrum, SPS models use libraries of observed or theoretical stellar spectra. Theoretical libraries are based on model atmospheres \citep[e.g.][]{CastelliKurucz2003}, and cover a wide parameter space. Empirical libraries such as MILES \citep{SanchezBlazquez2006} provide high-fidelity optical spectra but cover only the stellar parameters of nearby stars. Semi-empirical compilations such as BaSeL \citep{Lejeune1997} combine the two approaches. Cool, evolved stars (M, C and S giants and AGB stars) are poorly covered by all libraries, because molecular opacities, pulsation and circumstellar dust all affect their spectra \citep{Abdollahi23}. The SSP spectrum is the IMF-weighted sum of the library spectra along the isochrone, and the models of \citet{BruzualCharlot2003} and \citet{Maraston2005} are well-known examples.

\subsection{Dust models}
\begin{figure}[!t]
  \centering
   \includegraphics[width=0.4\textwidth]{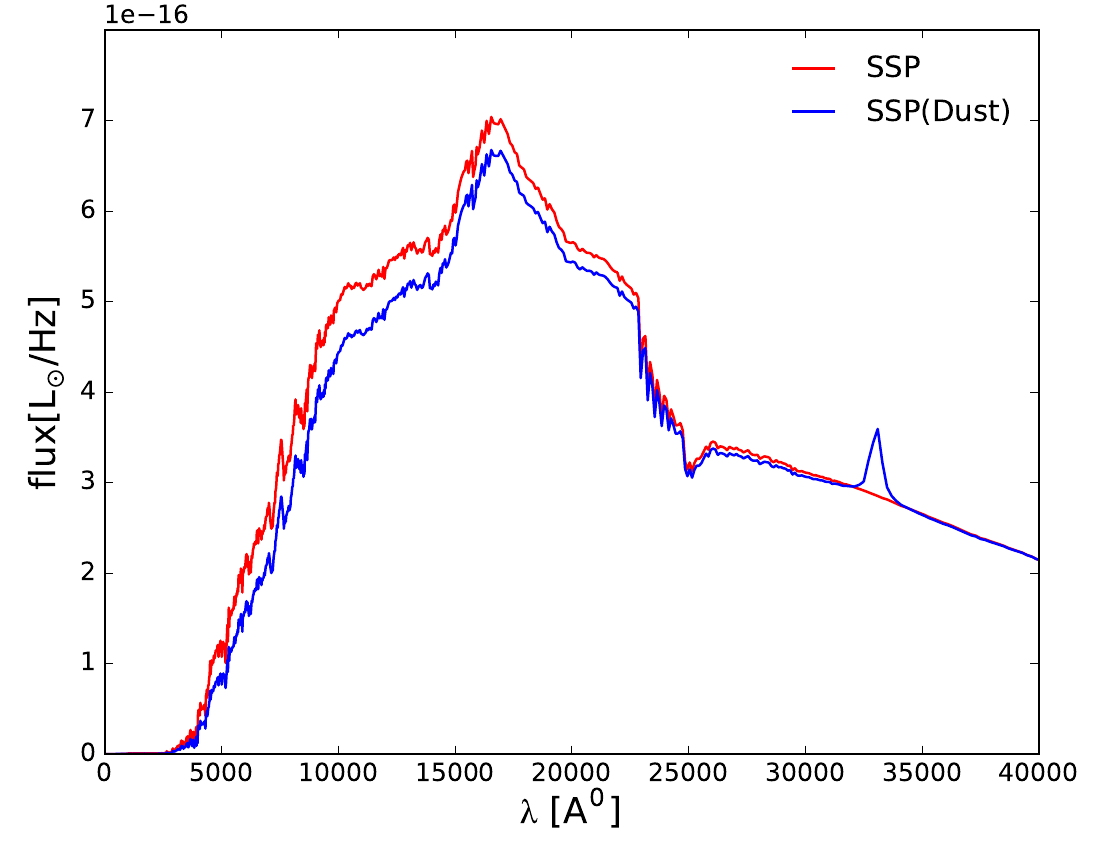}
  \caption{The effect of dust on the integrated luminosity. This figure illustrates the effect of dust on the integrated luminosity of a simple stellar population (SSP). As can be seen, dust attenuates the luminosity in the UV region while enhancing it in the IR region.}
  \label{fig:dust}
\end{figure}

Dust absorbs UV and optical photons and re-emits the absorbed energy in the IR, thereby modifying the SED over a wide wavelength range (see Fig.~\ref{fig:dust}). In stellar population synthesis, dust is commonly treated through three complementary approaches: \emph{attenuation laws}, which employ empirical or parametrized curves, such as the starburst attenuation law of \citet{Calzetti2000} and the two-component model of \citet{CharlotFall2000}, in which young stars experience additional attenuation from their birth clouds; \emph{grain models and radiative transfer in the ISM}, which account for grain size distributions and compositions, as described by \citet{MathisRumplNordsieck1977} and \citet{DraineLi2007}, as well as radiative-transfer codes such as GRASIL \citep{Silva1998}, which model the interaction between stars and dust within a galaxy's geometry; and \emph{energy-balance methods}, such as MAGPHYS \citep{daCunha2008}, which require the dust-emitted luminosity to balance the stellar luminosity absorbed by dust, thereby linking the UV--optical and IR portions of the SED. In addition to these approaches, a distinct local component is \emph{circumstellar dust}, which forms in the winds of evolved stars. Asymptotic giant branch (AGB) stars and red supergiants are major sources of dust in galaxies, and their circumstellar envelopes can be modelled using one-dimensional radiative-transfer codes such as \texttt{DUSTY} \citep{Ivezic97}, which solves the radiative-transfer problem by accounting for absorption, scattering, and thermal re-emission within a spherically symmetric dust shell. This provides the physical basis for the tools described in the following two sections.

\section{Method}
\label{sec:method}

To derive the circumstellar dust properties and mass-loss rates of evolved stars, we employ two complementary SED-fitting pipelines, \texttt{SEDust} and \texttt{MCM-SEDust} \citep{Mahani26,MahaniJavadi2025, Mahani_Farsi}. Both use radiative-transfer models generated with \texttt{DUSTY} \citep{Ivezic97,Elitzur01}, but they explore the parameter space differently: \texttt{SEDust} selects the best-fitting model from a precomputed grid through $\chi^2$ minimization, whereas \texttt{MCM-SEDust} performs on-the-fly radiative-transfer calculations within a Markov chain Monte Carlo (MCMC) framework, allowing continuous exploration of the parameter space and yielding posterior distributions.

\subsection{SED fitting with \texttt{DUSTY}}

The \texttt{SEDust} pipeline\footnote{\url{https://github.com/hmahani/SEDust}} automates the fitting of large catalogues of dust-enshrouded stars, for which individual trial-and-error fitting with \texttt{DUSTY} would be impractical \citep{Mahani25}. It uses a precomputed grid spanning stellar effective temperatures ($T_{\rm eff}=2000$--$3000$\,K, in steps of $100$\,K), inner dust temperatures ($T_{\rm d}=500$--$1500$\,K, in steps of $100$\,K), and circumstellar optical depths ($\tau=0.01$--$2.0$, with a step of $0.01$ below $\tau=0.5$ and $0.05$ above). Separate grids are constructed for carbon-rich (C-type) and oxygen-rich (M-type) stars. For each source, the observed fluxes are compared with the model predictions, and the best-fitting model is selected by minimizing
\begin{equation}
\chi^2 = \sum_{i=1}^{N_{\rm band}}
\frac{\left(F_i^{\rm obs}-sF_i^{\rm mod}\right)^2}{\sigma_i^2},
\label{eq:chi2}
\end{equation}
where $F_i^{\rm obs}$ and $F_i^{\rm mod}$ are the observed and model fluxes in the $i$th photometric band, $\sigma_i$ is the observational uncertainty, and $s$ is a scaling factor accounting for the luminosity normalization. The luminosity and mass-loss rate are then derived from the selected model.

To overcome the limitations of discrete grid-based fitting, in particular the inability to characterize parameter degeneracies and source-specific uncertainties, we developed \texttt{MCM-SEDust}\footnote{\url{https://github.com/AmirhosseinRabiee/MCM-SEDust}} as an extension of \texttt{SEDust}. At each sampling step it generates a new \texttt{DUSTY} input file from a user-defined template, runs the radiative-transfer calculation, and compares the resulting spectrum with the observed photometry. Three parameters are sampled: the stellar effective temperature ($T_{\rm eff}$), the dust temperature at the inner boundary of the shell ($T_{\rm d}$), and the optical depth at $2.2\,\mu$m ($\tau_K$). As in \texttt{SEDust}, the self-consistent radiatively driven wind solution of \texttt{DUSTY} is adopted, in which the density structure and terminal wind velocity are calculated internally.

An important advantage of \texttt{MCM-SEDust} is its flexibility in treating dust composition: grain properties, including chemical composition, size distribution, and optical constants, are specified through the input template rather than being restricted to precomputed grids. We consider separate carbon-rich and oxygen-rich templates. The carbon-rich composition consists of $85\%$ amorphous carbon and $15\%$ silicon carbide (SiC) \citep{Groenewegen06,Hanner88,Pegourie88}, while the oxygen-rich composition adopts astronomical silicates \citep{Draine84}. The same prescriptions were used in the NGC~6822 analysis, ensuring consistency when comparing the derived dust properties \citep{Mahani_Farsi}.

\subsection{Statistical fitting and comparison}

In \texttt{MCM-SEDust}, the observed photometry and model spectra are converted to $\lambda F_\lambda$ and compared in logarithmic flux space, with the model interpolated onto the observed wavelengths in log--log space. Fractional uncertainties are converted into logarithmic ones, $\sigma_i=(\sigma_F/F)/\ln 10$, and a fractional uncertainty of $15\%$ is adopted for measurements without published errors. Because the emergent \texttt{DUSTY} spectra have an internal flux normalization, the overall scaling is computed analytically at each step as the inverse-variance-weighted mean of the logarithmic residuals, rather than being treated as a fourth free parameter:
\begin{equation}
\Delta =
\frac{\displaystyle\sum_i
(\log F_i-\log M_i)/\sigma_i^2}
{\displaystyle\sum_i 1/\sigma_i^2},
\end{equation}
where $F_i$ and $M_i$ are the observed and model values of $\lambda F_\lambda$. The goodness of fit is then
\begin{equation}
\chi^2 =
\sum_i
\left[
\frac{\log F_i-\log M_i-\Delta}{\sigma_i}
\right]^2,
\end{equation}
with log-likelihood $\ln\mathcal{L}=-\chi^2/2$.

Uniform priors are adopted within $T_{\rm eff}=2000$--$4000$\,K, $T_{\rm d}=100$--$1400$\,K, and $\tau_K=0.01$--$2.0$; proposals outside these limits are rejected without running \texttt{DUSTY}, and models for which \texttt{DUSTY} fails to converge or exceeds the time limit are assigned $\ln\mathcal{L}=-\infty$. The posterior is sampled with the affine-invariant ensemble sampler \texttt{emcee} \citep{ForemanMackey13,Goodman10}, using 16 walkers initialized randomly within the priors, 128 burn-in steps and 1024 production steps for each source and dust composition. Sampling performance is assessed through the mean acceptance fraction and the integrated autocorrelation time ($\tau_{\rm int}$); we require a chain length of at least $\sim\!50\,\tau_{\rm int}$ and inspect the corner plots for multimodal distributions or solutions piled up at the prior boundaries. Refitting a representative source with a much longer chain gave consistent parameters within the uncertainties. Fits failing these checks are flagged and excluded.

\begin{figure*}[ht]
    \centering
    \includegraphics[width=0.45\textwidth]{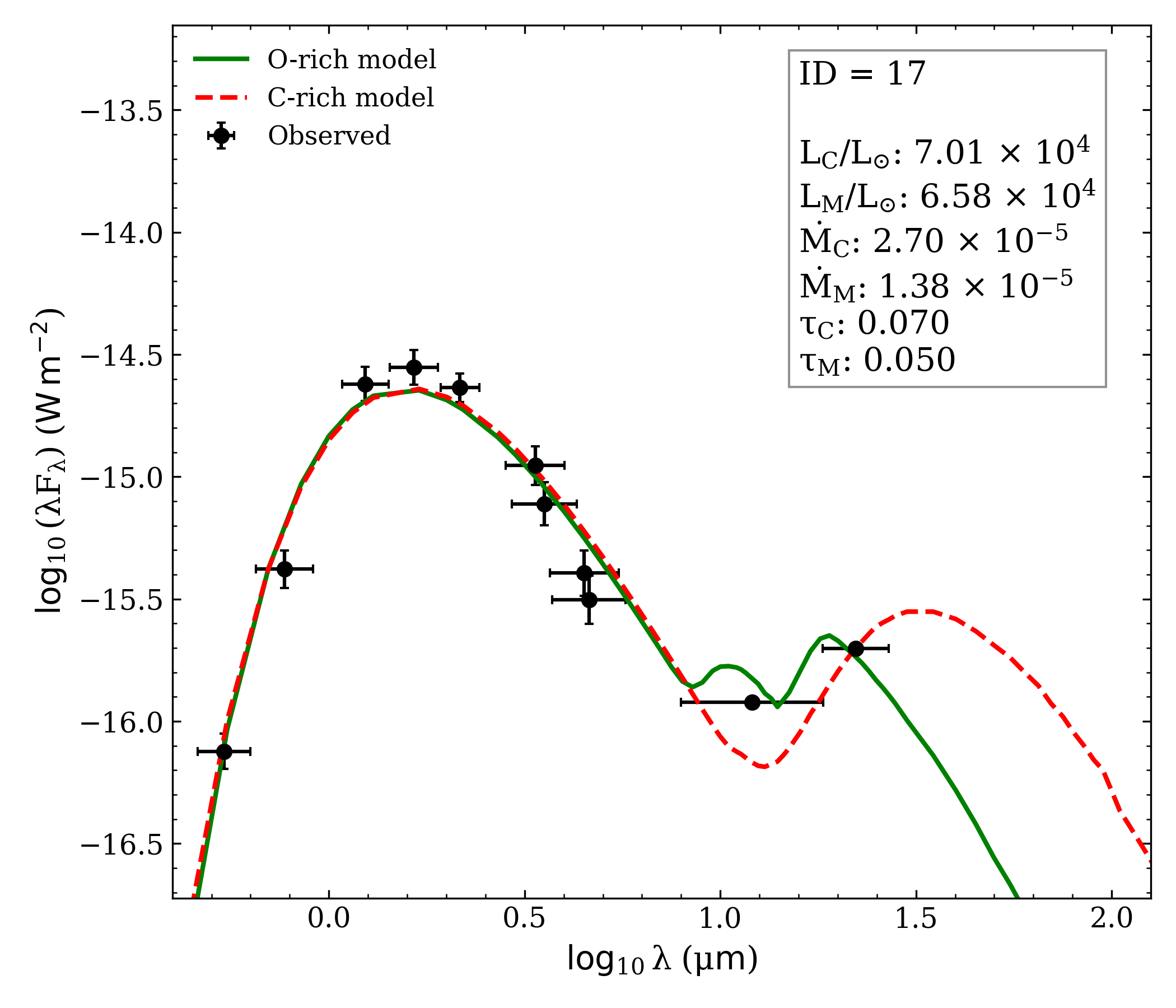}
    \includegraphics[width=0.45\textwidth]{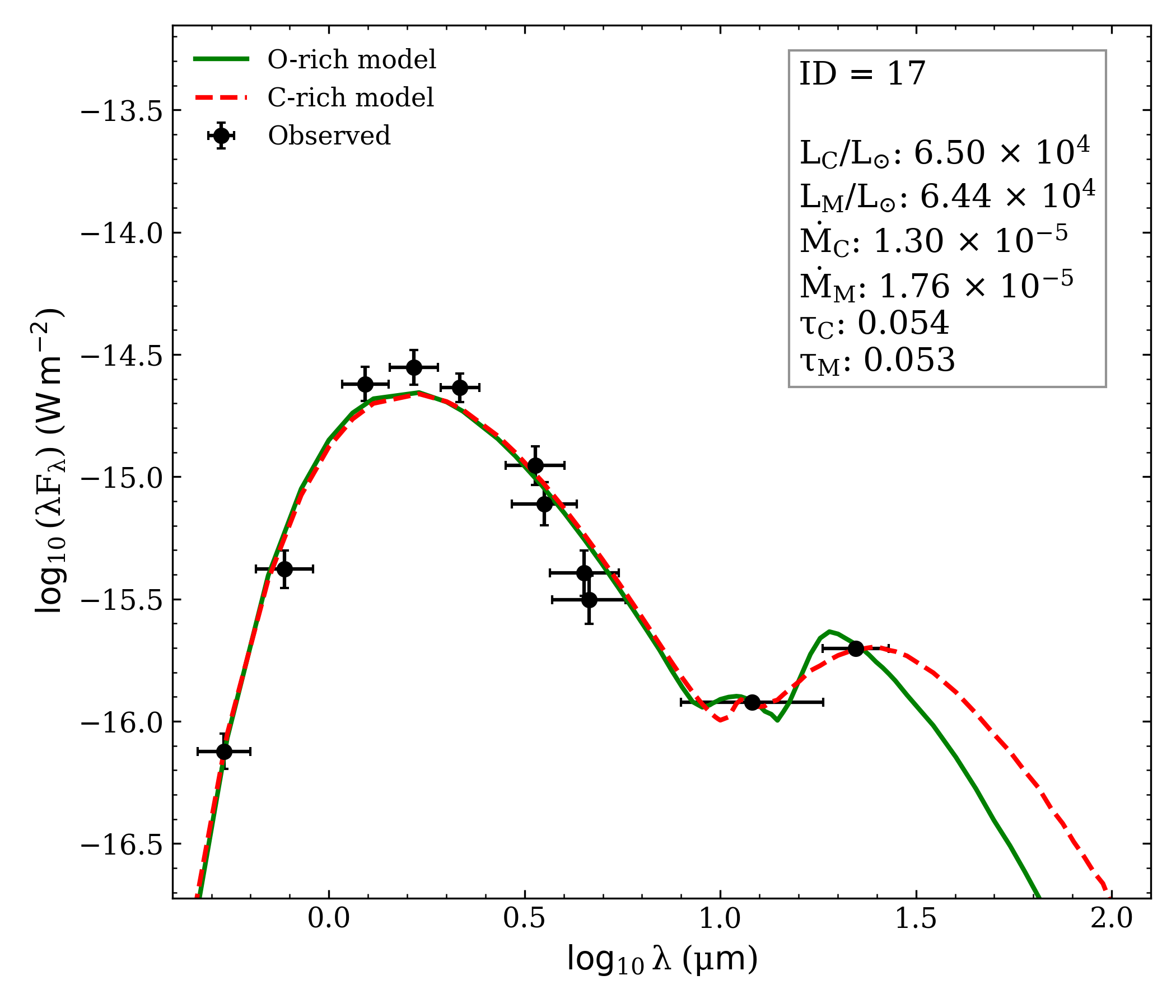}
    \caption{Comparison of the SED-fitting approaches implemented in
    \texttt{SEDust} and \texttt{MCM-SEDust}. The left panel shows an
    example best-fitting SED obtained using the precomputed
    \texttt{DUSTY} model grid in \texttt{SEDust}, while the right panel
    illustrates the corresponding MCMC-based fitting approach in
    \texttt{MCM-SEDust}, including the posterior distributions of the
    free parameters. Observed photometric measurements are shown as
    points, and the best-fitting model SEDs are represented by curves.}
    \label{fig:sedust}
\end{figure*}

Figure~\ref{fig:sedust} illustrates both pipelines for example AGB stars. For each source the posterior median is adopted as the representative value of each parameter and the 16th and 84th percentiles define the uncertainty interval; the luminosity and mass-loss rate are computed from a final \texttt{DUSTY} run at the posterior-median parameters. \texttt{MCM-SEDust} is computationally more expensive, since thousands of \texttt{DUSTY} runs may be needed per source. Both pipelines yield luminosities, optical depths, and mass-loss rates for individual evolved stars, which can be summed over a resolved population to estimate the total mass return and dust injection, providing observational constraints on the dust production assumed in SPS and galactic chemical evolution models.

\section{Results and Conclusions}
\label{sec:results}

To illustrate the differences between the two fitting approaches, we compare their results for an example IC 10 AGB star (ID~24472), using both carbon-rich and oxygen-rich circumstellar dust models (Table~\ref{tab:fit}). For \texttt{SEDust} the values correspond to the minimum-$\chi^2$ grid model, whereas for \texttt{MCM-SEDust} they are posterior medians with 16th--84th percentile intervals.

\begin{table}[ht]
\centering
\small
\caption{Best-fitting parameters for star ID~24472 obtained with \texttt{SEDust} (minimum-$\chi^2$ grid solution) and \texttt{MCM-SEDust} (posterior median and 68\% credible interval), for carbon-rich (C) and oxygen-rich (O) dust models.}
\label{tab:fit}
\begin{tabular}{lcccc}
\toprule
 & \multicolumn{2}{c}{\texttt{SEDust}} & \multicolumn{2}{c}{\texttt{MCM-SEDust}} \\
\cmidrule(lr){2-3}\cmidrule(lr){4-5}
Parameter & C-rich & O-rich & C-rich & O-rich \\
\midrule
$T_{\star}$ (K) & 2400 & 2400 & $2356^{+23}_{-15}$ & $2388^{+27}_{-18}$ \\
$T_{\rm d}$ (K) & 100 & 200 & $135^{+8}_{-7}$ & $172^{+10}_{-10}$ \\
$\tau$ & 0.070 & 0.050 & $0.054^{+0.006}_{-0.005}$ & $0.053^{+0.005}_{-0.004}$ \\
$L$ ($L_{\odot}$) & 70\,113 & 65\,775 & 65\,000 & 64\,438 \\
$\dot{M}$ ($M_{\odot}\,{\rm yr}^{-1}$) & $2.7\times10^{-5}$ & $1.4\times10^{-5}$ & $1.3\times10^{-5}$ & $1.8\times10^{-5}$ \\
\bottomrule
\end{tabular}
\end{table}

The two pipelines agree well on the effective temperature ($2400$\,K versus $2356^{+23}_{-15}$\,K for the carbon-rich model, and $2400$\,K versus $2388^{+27}_{-18}$\,K for the oxygen-rich model) and give comparable optical depths ($0.070$ versus $0.054^{+0.006}_{-0.005}$, and $0.050$ versus $0.053^{+0.005}_{-0.004}$). The inner dust temperatures differ more: the carbon-rich \texttt{SEDust} solution lies at the lower boundary of its grid ($100$\,K), whereas \texttt{MCM-SEDust} gives $135^{+8}_{-7}$\,K; for the oxygen-rich models the values are $200$\,K and $172^{+10}_{-10}$\,K. This shows how discrete sampling can restrict the best-fitting solution, while the MCMC approach explores the parameter space continuously.

The luminosities are consistent: for the carbon-rich models \texttt{MCM-SEDust} gives $65\,000\,L_\odot$, about $7\%$ below the \texttt{SEDust} value, and for the oxygen-rich models the two values differ by only $\sim\!2\%$ ($64\,438$ and $65\,775\,L_\odot$). The mass-loss rates differ more. The carbon-rich \texttt{MCM-SEDust} solution gives $\dot{M}=1.3\times10^{-5}\,M_\odot\,{\rm yr}^{-1}$ against $2.7\times10^{-5}\,M_\odot\,{\rm yr}^{-1}$ from \texttt{SEDust}, while for the oxygen-rich models the MCMC value ($1.8\times10^{-5}$) exceeds the grid value ($1.4\times10^{-5}\,M_\odot\,{\rm yr}^{-1}$). Similar optical depths and luminosities therefore do not guarantee similar mass-loss rates, which are sensitive to the adopted dust temperature and radiative-transfer solution.

We reviewed the main ingredients of SPS models and how dust shapes integrated SEDs, from the interstellar medium to circumstellar envelopes. AGB stars, long-period variables (LPVs) and red supergiants contribute substantially to the dust budget of galaxies, so an accurate treatment of circumstellar dust is essential for modelling their spectral properties and chemical evolution \citep{Javadi13_III, Abdollahi26, Javadi11, Boyer25, Boyer17, Goldman19}. We presented \texttt{SEDust}, an automated pipeline that fits precomputed \texttt{DUSTY} models to stellar SEDs and derives luminosities, optical depths and mass-loss rates, and which has been applied to NGC~147, NGC~185 and NGC~6822. We also introduced \texttt{MCM-SEDust}, which performs the radiative-transfer calculations within an MCMC framework and thus characterizes uncertainties and parameter degeneracies more directly than grid-based fitting. Together, the pipelines provide a framework for estimating the mass and dust returned by evolved stars in resolved galaxies, constraining the dust yields adopted in SPS and galactic chemical evolution models, particularly for the TP-AGB phase. Future work will extend the analysis to additional nearby galaxies and a wider range of dust compositions and grain properties, and will incorporate the resulting empirical constraints into SPS models.


\scriptsize
\bibliographystyle{ComBAO}
\nocite{*}
\bibliography{DUST_ComBAO}

\end{document}